 \documentclass[journal, comsoc]{IEEEtran}

\usepackage{gensymb}
\usepackage{cite}
\usepackage{amsmath,amssymb,amsfonts}
\usepackage{graphicx}
\usepackage{xcolor}
\usepackage{kotex}  
\usepackage{textcomp}
\usepackage{gensymb}
\usepackage{subcaption}

\usepackage{stfloats}
\usepackage{float}
\usepackage{wrapfig}
\usepackage{xcolor}
\usepackage{colortbl} 
\usepackage{color,soul}
\usepackage{ragged2e}
\usepackage{url}
\usepackage{verbatim}
\usepackage{graphicx}
\usepackage{cite}
\usepackage{multirow}
\usepackage{array,boldline, makecell,booktabs}
\usepackage{longtable} 
    \usepackage[group-separator={,},group-minimum-digits=4]{siunitx}

\begin{document}

\title{Towards 6G Evolution: Three Enhancements, Three Innovations, and Three Major Challenges}

\author{Rohit Singh, Aryan Kaushik,  Wonjae Shin, Marco Di Renzo, Vincenzo Sciancalepore, \\
Doohwan Lee,  Hirofumi Sasaki, Arman Shojaeifard, and Octavia A. Dobre \thanks{R. Singh is with the Department of Electronics and Communication Engineering, 
Dr B. R. Ambedkar National Institute of Technology Jalandhar, India, email: rohits@nitj.ac.in} \thanks{A. Kaushik is with the Department of Computing and Mathematics, Manchester Metropolitan University, UK (e-mail: a.kaushik@mmu.ac.uk).} \thanks{W. Shin is with the School of Electrical Engineering, Korea University, South Korea (e-mail: wjshin@korea.ac.kr). (\textit{Corresponding author: W. Shin})}

\thanks{M. Di Renzo is with Universit\'e Paris-Saclay, CNRS, CentraleSup\'elec, Laboratoire des Signaux et Syst\`emes, 3 Rue Joliot-Curie, 91192 Gif-sur-Yvette, France. (e-mail: marco.di-renzo@universite-paris-saclay.fr)} \thanks{V. Sciancalepore is with NEC Lab Heidelberg, Germany (e-mail: vincenzo.sciancalepore@neclab.eu).} \thanks{D. Lee and H. Sasaki are with the Network Innovation Laboratories, NTT  Corporation, Japan (e-mail: \{doohwan.lee, hirofumi.sasaki\}@ntt.com).} \thanks{A. Shojaeifard is with the InterDigital Europe, London, UK (e-mail: arman.shojaeifard@interdigital.com).}

\thanks{O. A. Dobre is with the Faculty of Engineering and Applied Science, Memorial University, Canada (e-mail: odobre@mun.ca).} }

\maketitle

%\vspace{-15cm}
\begin{abstract}
Over the past few decades, wireless communication has witnessed remarkable growth, experiencing several transformative changes. This article offers a comprehensive overview of wireless communication technologies, from the foundations to the recent wireless advances and future trends. Specifically, we take a neutral look at the state-of-the-art technologies for the fifth generation (5G) and ongoing evolutions towards 6G, while also reviewing the recommendations of the international mobile communication vision for 2030 (IMT-2030).  We first highlight specific features of IMT 2030, including three IMT-2020 extensions and three new innovations (Ubiquitous connectivity and integrating the new capabilities of sensing \& artificial intelligence with communication functionality). Based on key findings, we delve into three major challenges in implementing 6G, along with global standardization efforts. Besides, a proof of concept is provided by demonstrating terahertz (THz) signal transmission using orbital angular momentum (OAM) multiplexing, which is one of the potential candidates for 6G and beyond. To inspire further potential research, we conclude by identifying research opportunities and future visions on IMT-2030 recommendations.
\end{abstract}


\IEEEpeerreviewmaketitle

%\vspace{-3mm}

\section{{Introduction}}
\begin{figure}
    \begin{center}         %\vspace{-3mm}
    \includegraphics[width=0.82\linewidth]{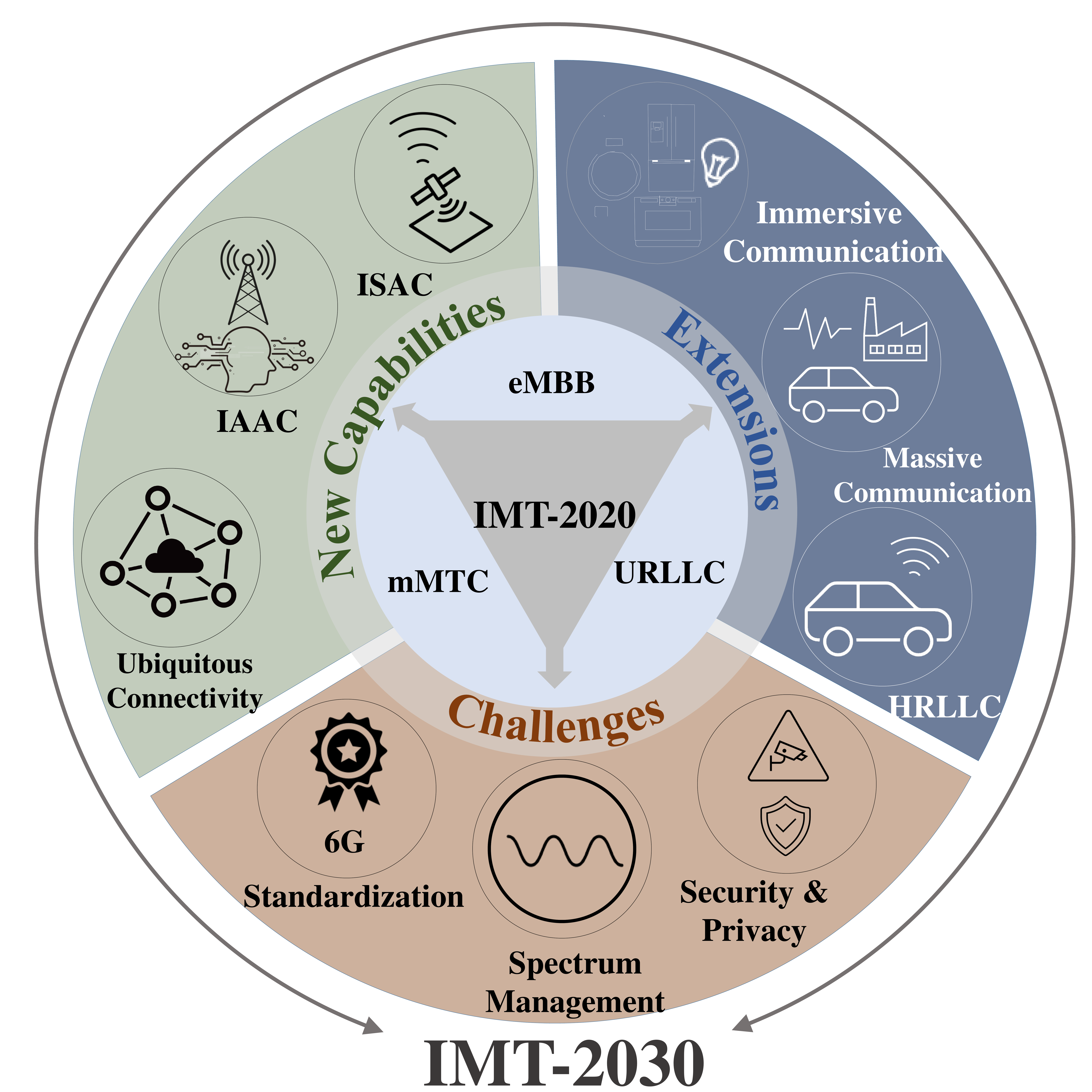}
    \end{center}
    \caption{An illustration of IMT 2030 capabilities, extensions, and associated challenges. \cite{imt}}
    \label{fig1}
\end{figure}

The cornerstone for the sixth generation (6G) has been set, with telecom agencies jointly working to advance its development. International mobile telecommunications (IMT) has already provided recommendations under the vision 2030, popularly known as IMT-2030 \cite{imt}. Learning from previous experiences, the new recommendations for 6G encompass the extensions of IMT-2020 capabilities and three new innovations, as depicted in Fig. \ref{fig1}. Specifically, these recommendations focus on better network flexibility and technology integration via making use of innovative features like artificial intelligence (AI), integrated sensing and communication (ISAC), and joint utilization of \textit{non-terrestrial networks} (NTNs) for ubiquitous connectivity. However, the proposed transformations bring new challenges in the form of effective spectral/energy utilization, the need for technology harmonization, and more importantly, security assurance under data-driven AI-assisted networks. Furthermore, there is no global consensus on the use of the terahertz (THz) spectrum for 6G, although researchers are trying to prove its effectiveness \cite{thz}.

Given the growing importance of 6G, this work aims to explore wireless trends and identify upcoming advancements, focusing on IMT 2030 recommendations. The goal is to highlight key lessons from previous generations, outline essential steps towards 6G, and identify technical considerations for more effective solutions. By reviewing the IMT 2030 recommendations, this work provides insights into the IMT 2020 extensions and three newly added capabilities, explaining how they complement the forthcoming network. {Besides, this work presents several essential aspects that remain unexplored in the existing works, mainly including a comprehensive technical exploration of 6G advancements and challenges, culminating in a proof of concept using THz transmission with orbital angular momentum (OAM) multiplexing. Further, this work addresses three major challenges that require urgent attention. Referring to the THz penetration issues and its unwavering potential, an enabling solution has been demonstrated, showcasing sub-THz transmission employing OAM multiplexing.} Finally, this work highlights research opportunities, offering insights into future directions based on the findings. {Specifically, the contributions of this work can be summarized as follows:}

\begin{itemize}
    \item {This article provides an in-depth technical analysis of the IMT-2030 hexagon, going beyond existing high-level illustrations to explore the underlying technological extensions and innovations in detail.}
    \item {In addition to tracing the historical evolution of wireless generations, this work highlights several key findings based on technical exploration, emphasizing the specific technological need at different levels for realizing the IMT-2030 vision.}
    \item {This study demonstrates a forward-looking proof of concept by showcasing THz signal transmission employing OAM multiplexing, thereby highlighting the feasibility and potential of advanced 6G technologies. }
\end{itemize}

\begin{figure*}
    \begin{center}         \includegraphics[width=0.81\linewidth]{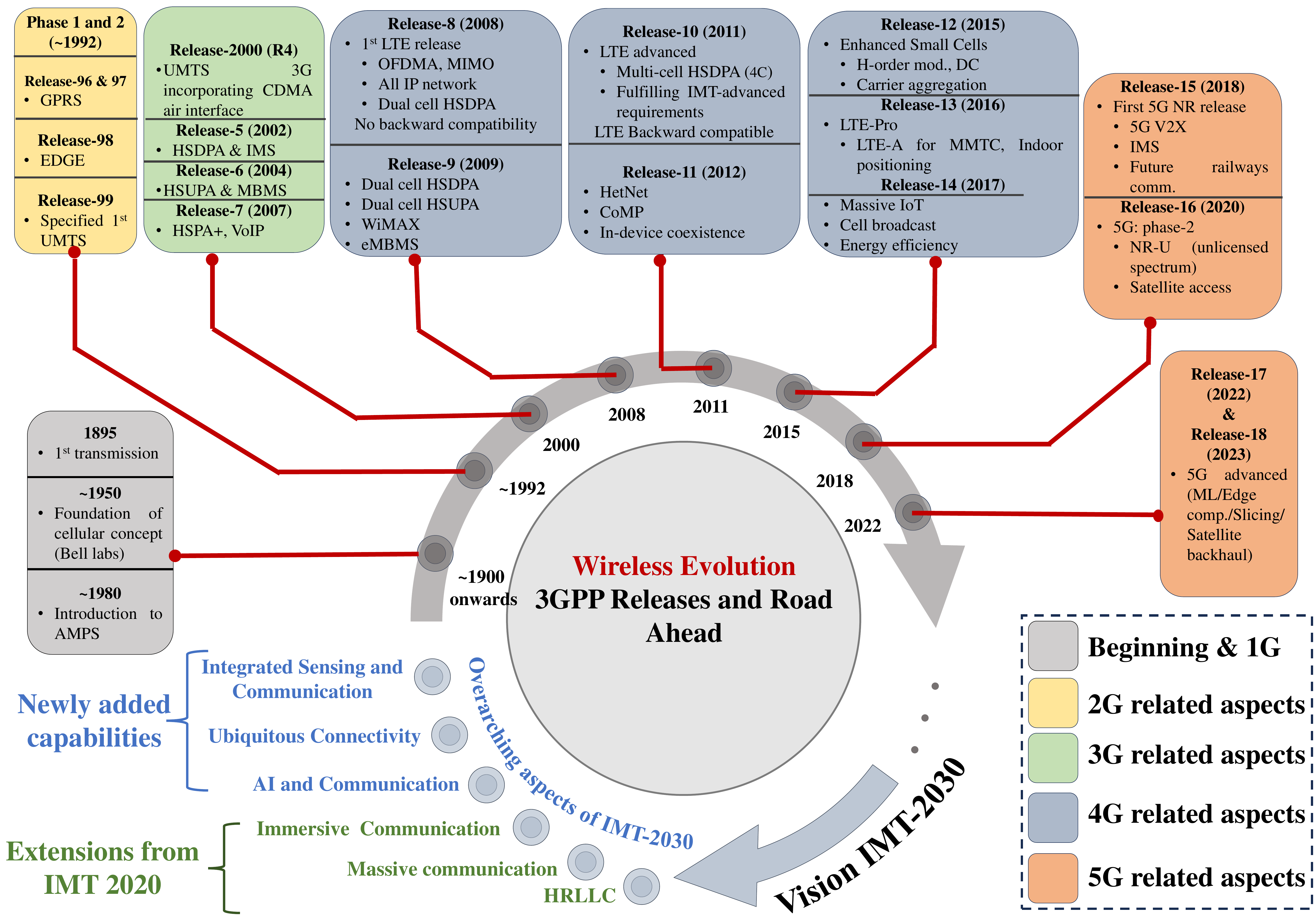}
    \end{center}
    \caption{{An illustration of wireless evolution timeline,  highlighting the periodic milestones associated with 3GPP releases.}}
    \label{fig2}
\end{figure*}

\subsection{Background and Motivation}
The wireless transmission of Morse code signals conducted in 1895 over a distance of 3.2 kilometers marks the inaugural instance of wireless transmission. Since then, wireless engineers have committed to accomplishing numerous milestones \cite{book}.\footnote{Telecommunication organizations, such as the International telecommunication union (ITU) and the third generation Partnership Project (3GPP) have played a crucial role in advancing wireless technologies. \cite{imt}} The early 2000s especially witnessed the arrival of third-generation (3G) cellular networks, bringing faster data transfer rates and the ability to support mobile Internet access. Moving into the late 2000s, the fourth generation (4G) emerged, substantially improving data speed, capacity, reliability, and latency. In the 2020s, the fifth generation (5G) took center stage as the latest wireless communication standard, promising ultra-fast data speeds, low latency, and the capability to connect a massive number of devices simultaneously. Currently, it is looking towards 2030 and beyond under the term \emph{IMT-2030 (6G)}. {Specifically, Release 21, expected to be finalized in the next year, signifies the formal initiation of normative 6G development and is anticipated to deliver the first official technical specifications for 6G, in alignment with IMT-2030 submission requirements.} As depicted in Fig. \ref{fig1}, IMT-2030 recommendations encompass the enhancement of three 5G pillars: ultra-reliable low latency communication (URLLC) in 5G extended to hyper-reliable low latency communication (HRLLC) in 6G, enhanced mobile broadband (eMBB) extended to immersive communication, and massive machine type communication (mMTC) extended to massive communication. Alongside this, new capabilities are being recommended, namely ISAC, ubiquitous connectivity, and integrated AI and communication (IAAC). These new capabilities are essential due to the following reasons: \textit{a)} users’ location now plays a critical role in advanced beamforming and interference avoidance where bringing together sensing capability holds potential for low complexity users sensing, \textit{b)} besides, recent network advances, especially core units rely on the AI assisted technologies for network virtualization and management, raising the need for IAAC extension, \textit{c)} finally, the advent of Internet of everything (IoE) is set to provide network support to billions of devices, making ubiquitous connectivity an essential step towards the forthcoming wireless era.

Previous wireless trends and 6G recommendations point to a more adaptive and interactive framework, offering vast possibilities but also presenting significant research challenges. By analyzing past trends and the IMT-2030 guidelines, the following sections explore key findings and outline the essential steps for 6G development. A detailed view is also provided on key technology enhancements and innovations, addressing three major challenges, depicted in Fig. \ref{fig1}, and summarised as \textit{i)} achieving global standardization and inter-technology harmony as 6G integrates both cellular and non-cellular wireless technologies and \textit{ii)} addressing data secrecy, as 6G will rely heavily on data-driven algorithms. Lastly, this work highlights research opportunities and future directions based on IMT-2030 recommendations.

\section{{Wireless Evolution and Trend Ahead}}
\subsection{{Evolution of Wireless Communication: From 1G to 6G}}
Wireless communication has undergone significant advancements, largely driven by advances in very large scale integrated (VLSI) circuits. Though its roots trace back to the early 19$^{th}$ century, the last fifty years have seen rapid development, with each decade marking the introduction of a new generation. Specifically, innovative technologies like multiple input multiple output (MIMO) and orthogonal frequency division multiplexing (OFDM) have emerged, providing more efficient transmission methods.
This section highlights the key advancements in each generation, highlighting the periodic release of 3GPP standards that shaped the wireless network, as depicted in Fig. \ref{fig2}. Further, this section discusses the lessons learned from the past evolution, drawing insights to guide future developments and trends.

\textit{The Beginning (1G):} The first generation of wireless communication emerged in the 1980s, marking the transition from landline telephony to mobile phones with the introduction of analog cellular networks. 1G systems provided basic voice communication but had limitations in terms of capacity, security, and signal quality. Specifically, the analog telephony system encountered numerous challenges, such as poor voice quality, limited coverage, low capacity, and frequent dropped calls, highlighting the need for a shift to digital technology.

\textit{Stepping into the Digital Era in 2G:} 2G, in the 1990s, experienced a shift from analog to digital communication, allowing for improved voice quality and the introduction of digital information, including text messaging. Specifically, two major 2G technologies, global system for mobile communications (GSM) and code division multiple access (CDMA), competed for dominance. The transition from analog to digital communication made it possible to offer a true data access experience via technologies like general packet radio service (GPRS), and enhanced data rates for GSM evolution (EDGE). Simultaneously, private players began developing their own standards, limiting global roaming and creating global standardization platforms like 3GPP for uniform standardization. 

\textit{Empowering CDMA in 3G:} In the early 2000s, the introduction of 3G brought about a leap in wireless technology with the foundation of global telecom organizations like 3GPP. 3G networks offered faster data transfer rates, enabling the support of mobile Internet access and more advanced services. Technologies like high-speed packet access (HSPA) and CDMA-2000 played a key role in providing higher data speeds, paving the way for a more interconnected and data-centric mobile experience.

\textit{Towards Effective Resource Utilization in 4G:} The late 2000s witnessed the advent of 4G, with faster data speeds, lower latency, and increased capacity compared to 3G. This era marked the rise of mobile broadband, enabling seamless video streaming, online gaming, and the widespread use of mobile applications. This decade also witnessed the foundation of several groundbreaking technologies, including MIMO, OFDM, all IP networks, and dual cell HSDPA (downlink), introduced in 3GPP Release-8 and Release-9. Moreover, a paradigm shift has been observed in subsequent 3GPP Releases, including Release-10, Release-12, and Release-14, which are responsible for the advent of wireless technologies like Multi-cell HSDPA and coordinated multi-point transmission (CoMP).

\textit{Utilizing Network Virtualization in 5G:} Launched near the 2020s, 5G is the latest evolution in wireless communication, creating the backbone for a diverse application under one umbrella. 5G utilizes advanced technologies such as mm-Wave spectrum, massive MIMO, and network slicing. On one hand, 5G brings faster and more reliable connections, making it suitable for critical applications in augmented reality (AR). On the other hand, 5G covers flexible network support for low data rate internet-of-things (IoT) use cases under RedCap \cite{red}.\footnote{5G reduced capability (RedCap) is a 5G standard that enables low-complexity, cost-effective, and battery-efficient devices to connect to 5G networks. It is designed to bridge the gap between 4G and 5G, catering to use cases that fall between the high-speed eMBB, ultra-reliable uRLLC, and low-throughput mMTC technologies.} The associated 3GPP releases are Release-15, Release-16, and beyond, solely focusing on high-end applications, including vehicular/railway communication, satellite access, edge computations, etc.

{\textit{6G Expectations:} Building upon the foundational achievements of previous wireless generations and in alignment with IMT 2030 recommendation, it is evident that 6G revolves around achieving ultra-high-speed, low-latency, intelligent, and sustainable communication systems that can seamlessly sustain inter-technology integration. Such as integrated communication and sensing (ISAC), enhanced security frameworks including post-quantum cryptography, and the use of higher-frequency bands for ultra-dense data transfer.  Besides, emerging use cases raise the need for massive device connectivity (supporting up to 10⁷ devices/km²). More importantly, AI will play a key role in the network architecture design and management, enabling real-time optimization and automation. These technical expectations are strategically aligned with the 3GPP Release 20 and Release 21 timelines, evolving in a phase-wise manner. For instance, Release 20 (2024–2025) is considered the final stage of 5G-Advanced, focusing on enhancements in AI/ML, XR, positioning, and energy efficiency. Release 21 (2025–2027), on the other hand, marks the beginning of normative 6G work and is expected to produce the first formal technical specifications for 6G. This release will introduce emerging 6G technologies, including the early exploration of AI-native networking architectures, and enable support for next-generation use cases such as digital twins, immersive communication, and metaverse-enabling services. }

\subsection{{Lessons Learned and Trend Ahead}}
The above discussion confirms that wireless technology has advanced significantly, fueled by parallel developments in VLSI technologies. Further, it can be inferred from the above discussion that the wireless network has evolved in all aspects: \textit{a)} a dramatic increase in achievable data rate through continuous improvements in spectral efficiency and multiple access techniques; \textit{b)} enhanced coverage, reaching remote areas and crossing walls, made possible by advanced channel estimation and processing; \textit{c)} lightning-fast speeds, enabled by expanding spectrum and ongoing efforts toward the high-frequency spectrum; and d) support for diverse applications, from basic voice transmission in 1G to text messaging in 2G, multimedia in 3G/4G, and now critical and non-critical applications in the flexible 5G network.

Indeed, wireless researchers explored every possible aspect to maximize the potential of wireless networks. {According to Shannon's capacity theorem, summarised as $NB\log_2{(1+SINR)}$. Here, $N$ denotes the number of antenna pairs at the transceiver, $B$ is the channel bandwidth, and $SINR$ (an acronym for signal-to-interference-and-noise ratio) reflects channel conditions.} Let's see how things evolved from Shannon's viewpoint: \textit{a)} shifting to a higher frequency spectrum has significantly increased per-user bandwidth, from tens of kHz to hundreds of MHz, with carrier aggregation further expanding assigned bandwidth to a few GHz. Efforts are now focused on increasing transmission frequencies to the THz band, as demonstrated by a proof of concept in Section \ref{sec5}, showcasing over OAM modulation; \textit{b)} higher-order MIMO provides multiple parallel data paths, scaling data rates up to 64 MIMO streams, though this increases channel training and processing complexity. Researchers are trying to multiply this number through extensive massive MIMO; \textit{c)} reconfigurable environments, utilizing intelligent reflecting surfaces (IRS), are envisioned to enhance signal strength. Unlike traditional methods, where channel impairments were considered uncontrollable, the effective use of intelligent environments could further improve Shannon’s capacity.

\begin{table*}
   \centering
    \caption{An Overview 6G Technologies and Application Support.}
\begin{tabular}{|p{3 cm}|  p{7.5 cm} | p{6 cm}|} 
 \hline
\textbf{Technology} & \centering \textbf{Strategy}    &  \textbf{Key Technology and Aspects}\\  
 \hline
\vspace{-25mm}  ISAC and HRLLC&  
   \centering
    \includegraphics[width=0.85\linewidth]{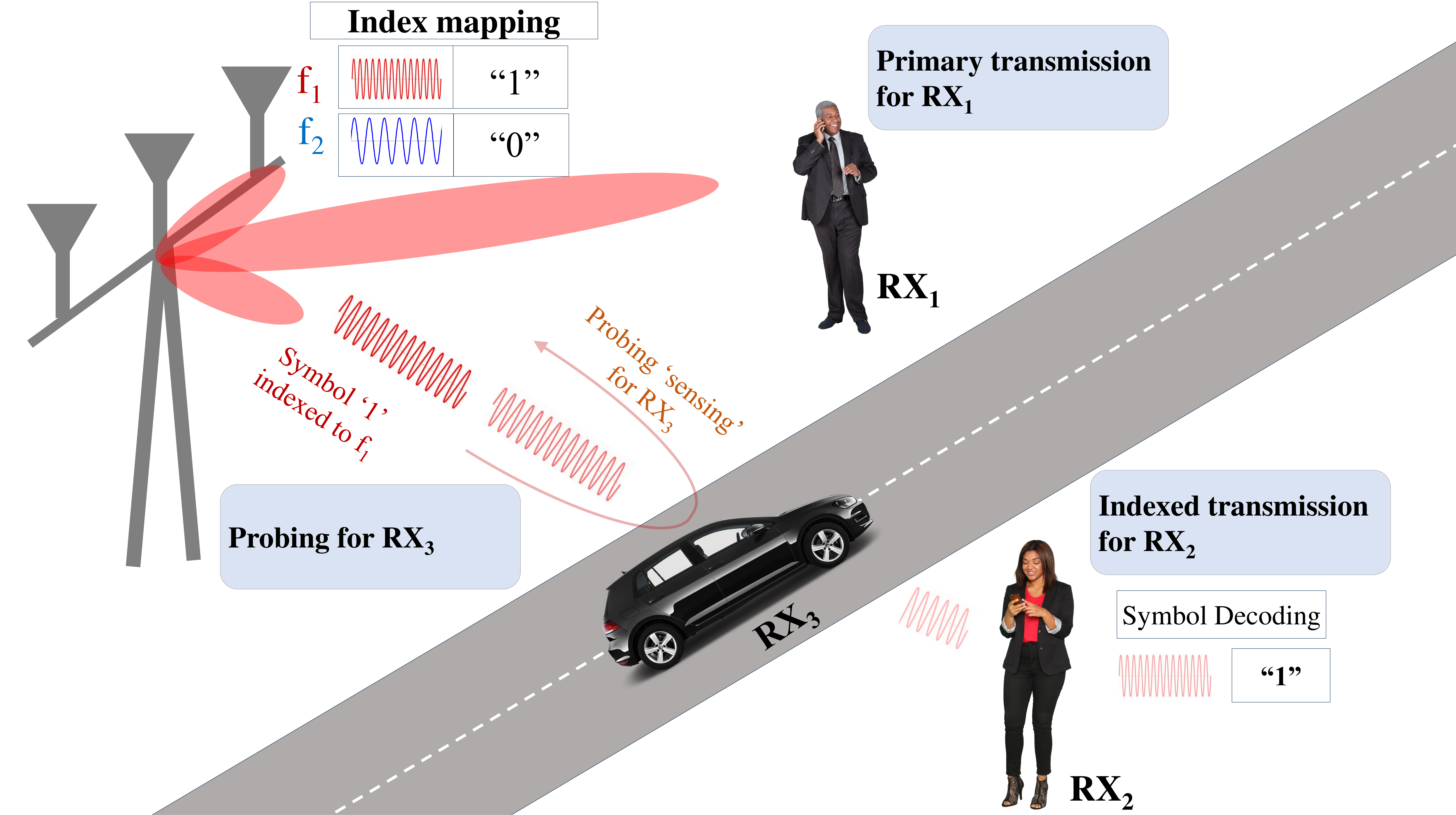}
    \captionsetup{labelformat=empty}
  & \vspace{-25mm} 
      \underline{Enhanced Channel Training and Beamforming}
  \begin{itemize}
    \item ISAC assisted passive channel estimation
    \item Predictive beamforming via AoA Estimation
\end{itemize} \\ 
  \hline
\vspace{-25mm}  IAAC and Immersive Communication&       
\centering
\includegraphics[width=0.9\linewidth]{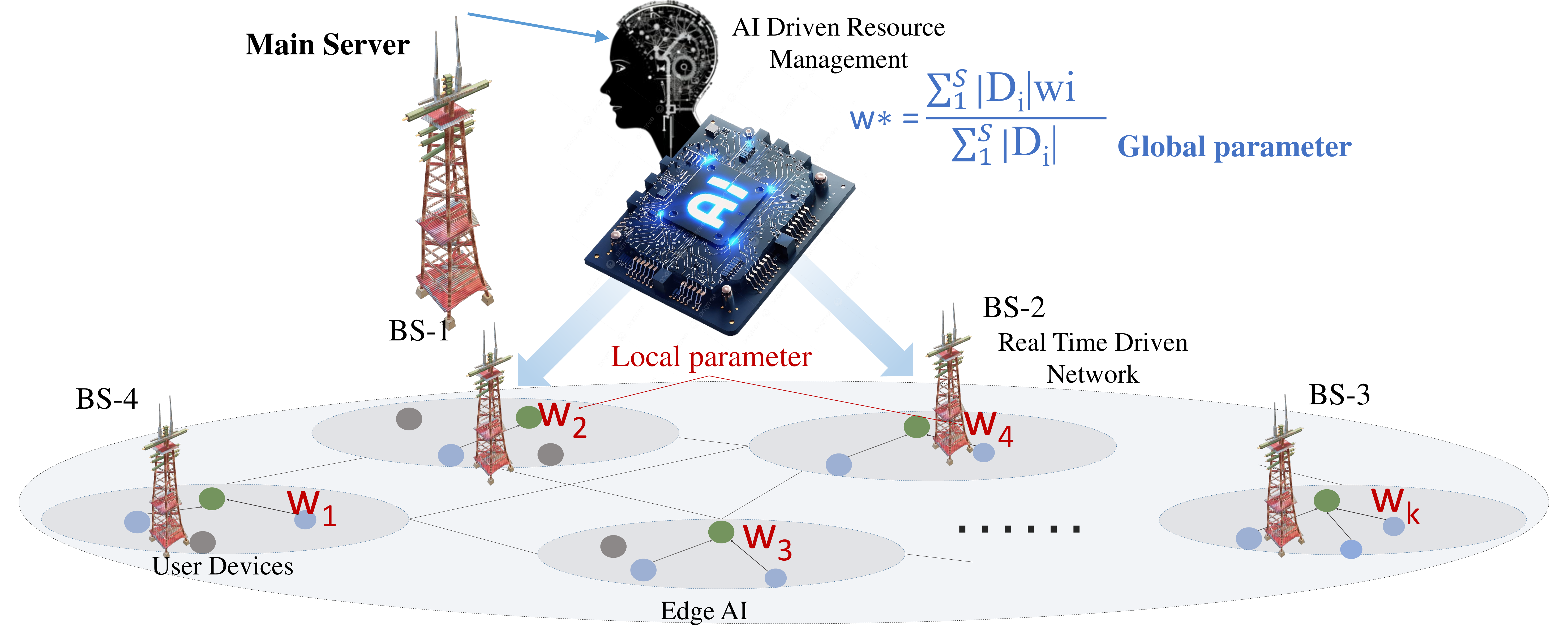}
    \captionsetup{labelformat=empty}
 & \vspace{-25mm} 
 \underline{Enabling AI for Network Virtualization}
 \begin{itemize}
    \item AI-assisted virtualization leads to seamless NFV. 
    \item Further, it enables auto-scaling based on network load, handling potential threats and faults. 
    \end{itemize}\\
  \hline
 \vspace{-25mm} {Ubiquitous Connectivity and Massive Communication} &  
 \centering
 \includegraphics[width=0.9\linewidth]{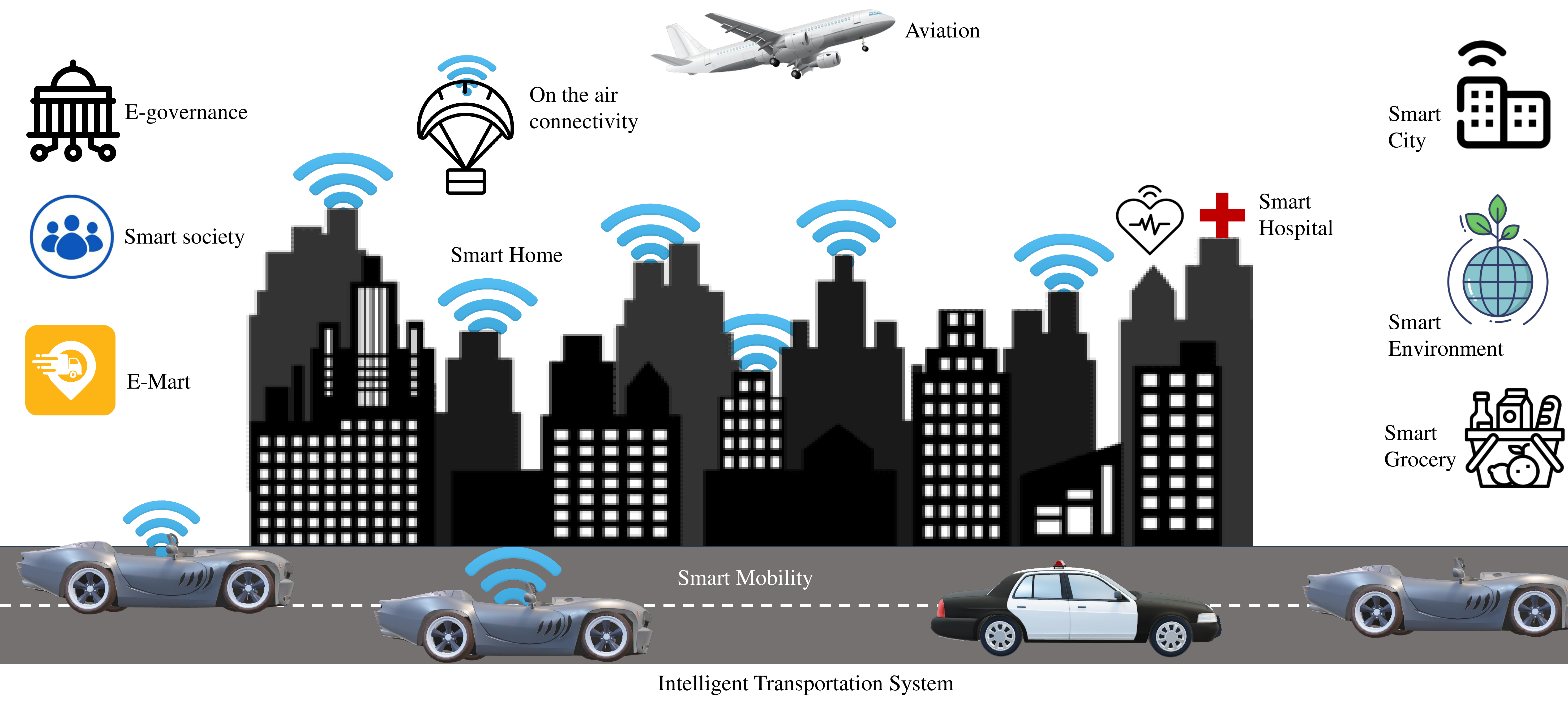}
    \captionsetup{labelformat=empty}
& \vspace{-25mm}
\underline{Ubiquitous Connectivity under Smart Environment}
\begin{itemize}
    \item A more precise indoor and outdoor positioning.
    \item Smart (mobile, wireless, service) devices and environments for a seamless man-machine interaction.
    \end{itemize} \\
 \hline

\end{tabular}
\label{tab1}
\end{table*}

\section{{Pathway to 6G: IMT 2030 Enhancements and Innovations}}
The previous section discussed the evolution of wireless networks to date and highlighted future trends. While significant advancements have already been made, the focus is now shifting towards a more flexible and unified framework that accommodates a wide range of use cases. Table I provides a comprehensive overview of key 6G technologies. These are: \textit{a)} the first row demonstrates ISAC-assisted sensing for advanced beam prediction, utilizing transmitted signals to estimate the Angle of Arrival (AoA) through the reflected signal as a probing mechanism, \textit{b)} the next row highlights AI-driven virtualization for seamless network function virtualization (NFV), enabling management for network load while effectively handling potential threats and faults, and \textit{c)} the last row demonstrates ubiquitous connectivity within a smart environment, enabling more accurate indoor and outdoor positioning. Moreover, this section delves into the IMT 2030 hexagon, including three enhancements and three innovations. Moreover, this section summarises key research findings and essential steps towards the development of the forthcoming network.

\subsection{IMT 2020 Extensions: A Closer View}
IMT 2030 networks represent an evolution from its predecessor, IMT 2020, which encompasses the extension of the key pillars, including URLLC, eMBB, and mMTC \cite{6gb}.

\subsubsection{Enhanced URLLC (URLLC+)}
URLLC, a core feature of 5G, ensures ultra-reliable communication with minimal latency, supporting applications that require immediate responsiveness. Building on this, the transition from URLLC to HRLLC aims to enhance reliability and responsiveness further, making it even more suitable for hyper-reliable applications. Achieving this requires shifting of processing capabilities closer to the end-user to reduce delays. Consequently, techniques like edge computing, combined with modern learning tools at the distributed edge, help minimize transmission delays and enhance data security by keeping computations closer to the source. Moreover, research efforts are being made to use the THz band for extremely high data rates, ensuring reduced reliability.

\subsubsection{Redefining eMBB (eMBB+)}
The transition from eMBB in 5G to immersive communication in 6G represents a focus on higher data rates, increased network capacity, and enhanced user experiences. Specifically, immersive communication in 6G aims to deliver more interactive and engaging experiences. From a high-end application view, this shift can be enabled through the adoption of technologies such as extremely large-scale MIMO (XL-MIMO) and cell-free MIMO (CF-MIMO) \cite{xmm}. While XL-MIMO provides extensive spatial degrees of freedom (DoF), CF-MIMO ensures seamless connectivity beyond traditional cell boundaries, improving handover and enhancing user performance at the cell edge. Nonetheless, the focus of 6G is not limited to high-end applications; features like network slicing and extended RedCap strive to provide a seamless integration of low data rate applications.

\subsubsection{Beyond mMTC (mMTC+)} 
The shift from mMTC to massive communication marks a significant advancement in telecommunications. This transition aims to form a versatile and interconnected communication ecosystem, enabling it to support diverse users' requirements.
While mMTC in 5G focuses on connecting a vast number of devices, massive communication focuses on future developments, such as satellite integration, which potentially transform mobile and IoT networks.\footnote{Notably, 3GPP has officially launched research into integrating satellite communications with 5G New Radio (NR) technologies, known as NTN. \cite{isc}}

\textbf{Key Findings and Essential Steps:} 
Overall, IMT 2020 extensions ensure better coverage, speed, and scalability, opening the doors for manifold benefits from household and industry perspectives. For instance, HRLLC in manufacturing is intended to offer seamless communication between robotic systems and facilitate control of production processes. On the other hand, HRLLC makes it possible to support remote surgeries and medical procedures where latency is paramount. Shifting to immersive communication is more than an incremental improvement over its predecessors with a broader exploration of communication possibilities in 6G. Similarly, massive communication aims to form a versatile and interconnected communication ecosystem, enabling it to support diverse users’ requirements.

Indeed, it would be challenging to ensure the required degree of reliability. For instance, edge computing undoubtedly brings manifold advantages, yet it supports limited computation, storage, and communication resources compared to traditional cloud solutions and heterogeneous infrastructures. Similarly, deploying a large number of antennas brings more parallel channels, but it comes at the cost of vast channel estimation challenges over such a large array, especially in high mobility scenarios where the coherence window remains significantly small. Nonetheless, global efforts are being made to overcome these limitations through complementary network and wireless technologies. For instance, study item (SI) from Release 15 and beyond involves the identification of potential NTN scenarios, architectures, fundamental NTN issues, and corresponding solutions. Further, it is to support 5G NR’s basic features through both regenerative and transparent satellite systems.\footnote{Besides, organizations like the ITU are working on spectrum harmonization of NTN since satellites are likely to cover beyond the physical boundaries.}

\subsection{6G Innovations: Newly Added Capabilities}
In addition to the extensions of 5G technologies, IMT 2030 encompasses three new capabilities, including IAAC, Ubiquitous connectivity, and ISAC.

\subsubsection{Ubiquitous Connectivity}
Ubiquitous connectivity envisions a network infrastructure well beyond the traditional boundaries \cite{ubc}, ensuring that the users experience a seamless connection regardless their location or mobility. Nonetheless, the ubiquitous deployment can only be brought through the seamless connection of sensors and devices, offering innovative solutions for modern problems like agriculture experience via precision farming techniques. Besides, such interactions lead to a heterogeneous scenario, which can bring through techniques like cell-free architectures, including multiple access points that collaborate to enhance reliability and reduce interference.

\subsubsection{Integrated Sensing and Communication}
ISAC is another transformative capability which brings together a seamless fusion of advanced sensing technologies and robust communication systems via the real-time gathering, processing, and exchange of data \cite{hrl}. Specifically, the synergy of sensing and communication prepares a backbone for the applications that demand the integration of sensor data, improved resource utilization, and a more interconnected and responsive ecosystem. For instance, ISAC-enabled approaches are utilized for users' localization for simultaneous channel estimation and beamforming, leading to enhanced communication performance, as shown in Table \ref{tab1}.\footnote{The coexistence of communication-sensing paves the way for several industrial opportunities, including supporting novel industrial features, e.g., production lines, equipment monitoring, and logistics coordination.} 

\subsubsection{IAAC}
The amalgamation of AI and communication for 6G signifies a groundbreaking step toward an intelligent, adaptive, and secure wireless framework \cite{aic}. Incorporating AI-driven predictive algorithms facilitates more optimized resource management, predictive bandwidth demands, and optimizing spectrum usage, leading to a more reliable communication infrastructure. Moreover, the integration extends to the network's edge, enabling intelligent edge computing for faster response times in applications such as AR and autonomous systems. Furthermore, AI enhances network security by proactively detecting and responding to potential threats, fortifying the communication infrastructure against cyber attacks. 

\textbf{Key Findings and Essential Steps}
Overall, new capabilities in 6G focus on network flexibility and technology integration. Whereas ubiquitous Connectivity extends network infrastructure beyond traditional limits, the coexistence of communication and sensing in ISAC improves communication performance joint sensing and communication. Besides, integrating AI into communication frameworks provides more intelligent and adaptive systems. Accordingly, the forthcoming network will benefit from advanced sensing technologies such as user localization and simultaneous channel estimation. Moreover, IAAC marks a pivotal advancement, providing network security by proactively detecting threats and supporting intelligent edge computing for faster response times in applications such as AR and autonomous systems. 

Indeed, the implementation of such a network requires collaborations among distinct technologies. Similar to other wireless frameworks, reduced interference will be the key to seamless device integration to enable smooth communication in diverse settings, where ISAC fusion can do wonders with precise user localization and beamforming. Moreover, AI-driven algorithms must be incorporated to optimize resource management and spectrum usage, making networks more efficient. Overall, ensuring such collaboration across technologies requires solutions to such challenges, standardizing protocols through organizations, and fostering multi-stakeholder partnerships to align efforts and goals. The next section primarily focuses on 6G implementation challenges and efforts.

\section{{Pathway to 6G: Associated Challenges and Efforts}}\label{sec4}
\begin{figure}
    \begin{center}         \includegraphics[width=0.82\linewidth]{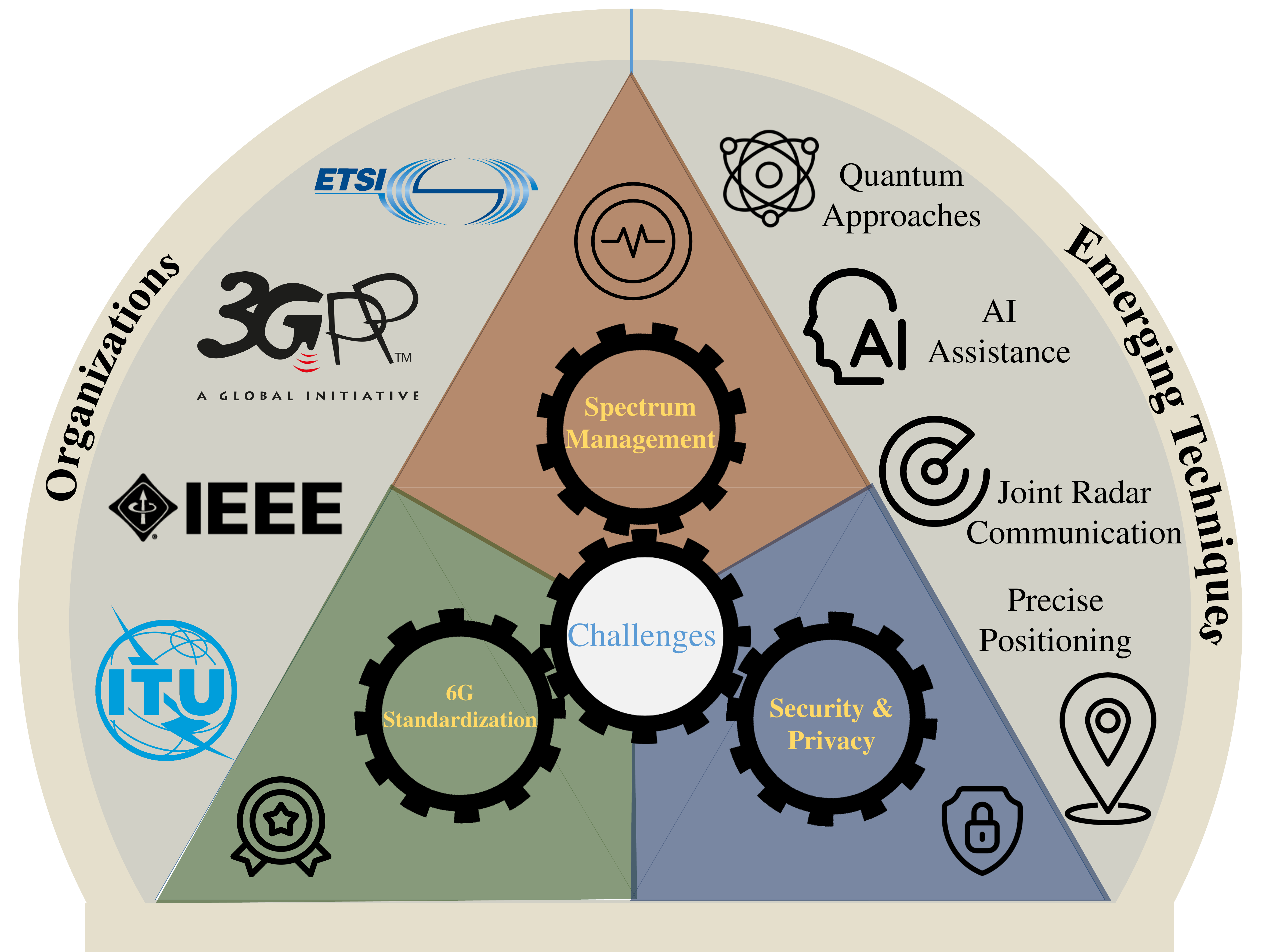}
    \end{center}
    \caption{{Key 6G challenges and corresponding efforts by organizations and emerging technologies
    %An illustration of associated challenges and ongoing efforts.
    }}
    \label{fig5}
\end{figure}

{Referring to Fig. \ref{fig1}, the paper outlines three key innovations, three notable extensions, and three major challenges. This section delves into these challenges and highlights ongoing efforts to address them, as illustrated in Fig. \ref{fig5}.}  

\subsection{{Associated Challenges: An Insight}}
{As discussed, 6G aims to unify a broad range of use cases supported by diverse technologies within a unified framework. Achieving this vision necessitates a globally coordinated standardization effort to ensure seamless integration, efficient spectrum planning and management, and a transparent strategy
to strengthen security and privacy. }

{\textit{6G Standardization:} Standardization is one of the critical pillars in shaping the global 6G landscape, though it presents unique challenges that must be addressed to realize a unified and future-ready framework; \textit{a)} one of the challenges lies in establishing a consensus on key performance requirements, use cases, and architecture models, \textit{b)} another challenge arise in the form of AI-enabled adoption as  6G aims to utilize AI at the core of network operations, demanding standards for training and deployment. Despite AI holds significant potential to address the associated issues, incorporating AI across the protocol stack poses another critical challenge, \textit{c)} moreover, the divergence in regional and organizational priorities, i.e., among standard bodies like 3GPP, ETSI, etc., may impose barriers to achieving global harmonization, and \textit{d)} finally, the inclusion of recently emerging technologies, e.g., quantum communication, IRS, etc., may require redefining channel models and network functionalities, making the standardization process technically intensive. }

{\textit{Spectrum Planning and Management:}  This emerges as another challenge due to the pursuit of additional bands, increasing the technical, regulatory, and strategic hurdles, for instance; \textit{a)} while a few research studies find THz bands highly useful to support terabit-per-second (Tbps) data rates \cite{thz}, global regulatory consensus is still lacking, \textit{b)} beside, operating in the higher perspective range brings additional propagation challenges, requiring innovations in adaptive modulation and channel estimation techniques, \textit{c)} yet another challenge occurs due to the coexistence with existing systems as backward compatibility with 5G and earlier systems must be carefully managed to ensure smooth network evolution.}

{\textit{Security and privacy:} As 6G systems are envisioned to support mission-critical and real-time applications, securing the underlying AI models becomes a critical aspect of 6G cybersecurity. Threats such as data poisoning and adversarial machine learning can significantly undermine the integrity of the autonomous decision-making process. Although several innovative solutions are emerging to address these concerns, they bring new challenges; a) ISAC, while promising, may expose sensitive environmental or user data if not carefully managed, leading to potential privacy breaches, b) likewise, quantum computing presents a dual-edged scenario as it enables ultra-secure communication through quantum key distribution but simultaneously threatens current cryptographic systems in the long term \cite{quant}, c) furthermore, the convergence of communication, computation, and storage at the network edge introduces edge-centric security risks, as data is processed and retained closer to end users. Collectively, the key challenge is to design lightweight, scalable security protocols that can operate effectively across highly heterogeneous and resource-constrained devices.
}

{\textbf{Key Findings and Essential Steps:} In light of the aforementioned challenges, it is clear that while 6G presents several opportunities, it also accompanies several challenges that require a coordinated and forward-looking strategy. Although IMT-2030 outlines key expectations and formal groundwork for 3GPP Release 21, the scope of 6G necessitates ongoing technological exploration and innovation. Crucially, standardization efforts must be accelerated by early-stage collaborative research and active policy engagement to ensure global coherence and alignment. Simultaneously, the effective spectrum utilization demands advancements in radio technologies, dynamic spectrum sharing mechanisms, and adaptable licensing models, necessitating harmonized regulatory frameworks for scalable and reliable deployment. Furthermore, security continues to be a foundational pillar for 6G, particularly with the integration of AI and sensing technologies, which pose new challenges in ensuring transparency, real-time responsiveness, and data privacy. Addressing these concerns requires the development of advanced encryption methods and privacy-preserving mechanisms that can secure both communication and sensory data. }

\section{{Proof of Concept, Future Prospects, and Research Directions}}\label{sec5}
Among the key wireless advancements highlighted in previous sections, the continuously growing demand for higher data rates remains critical, emphasizing the need to utilize the THz spectrum. Acknowledging the severe attenuation challenges associated with THz frequencies, substantial research efforts are underway to address and mitigate these issues. A promising technology is OAM multiplexing, which efficiently utilizes sub-THz communications. As a proof of concept, this section demonstrates the transmission using an OAM multiplexing performed by Sasaki et al. \cite{sas}. {The successful implementation of sub-THz OAM multiplexing, one of the most technically challenging 6G technologies, demonstrates the feasibility of other emerging innovations, such as ISAC, IAAC, and ubiquitous connectivity.} This section summarizes the directions of future research and parallel activities.

\subsection{Proof of Concept: Sub-THz OAM transmission}
\begin{figure}
    \begin{center}         \includegraphics[width=1\linewidth]{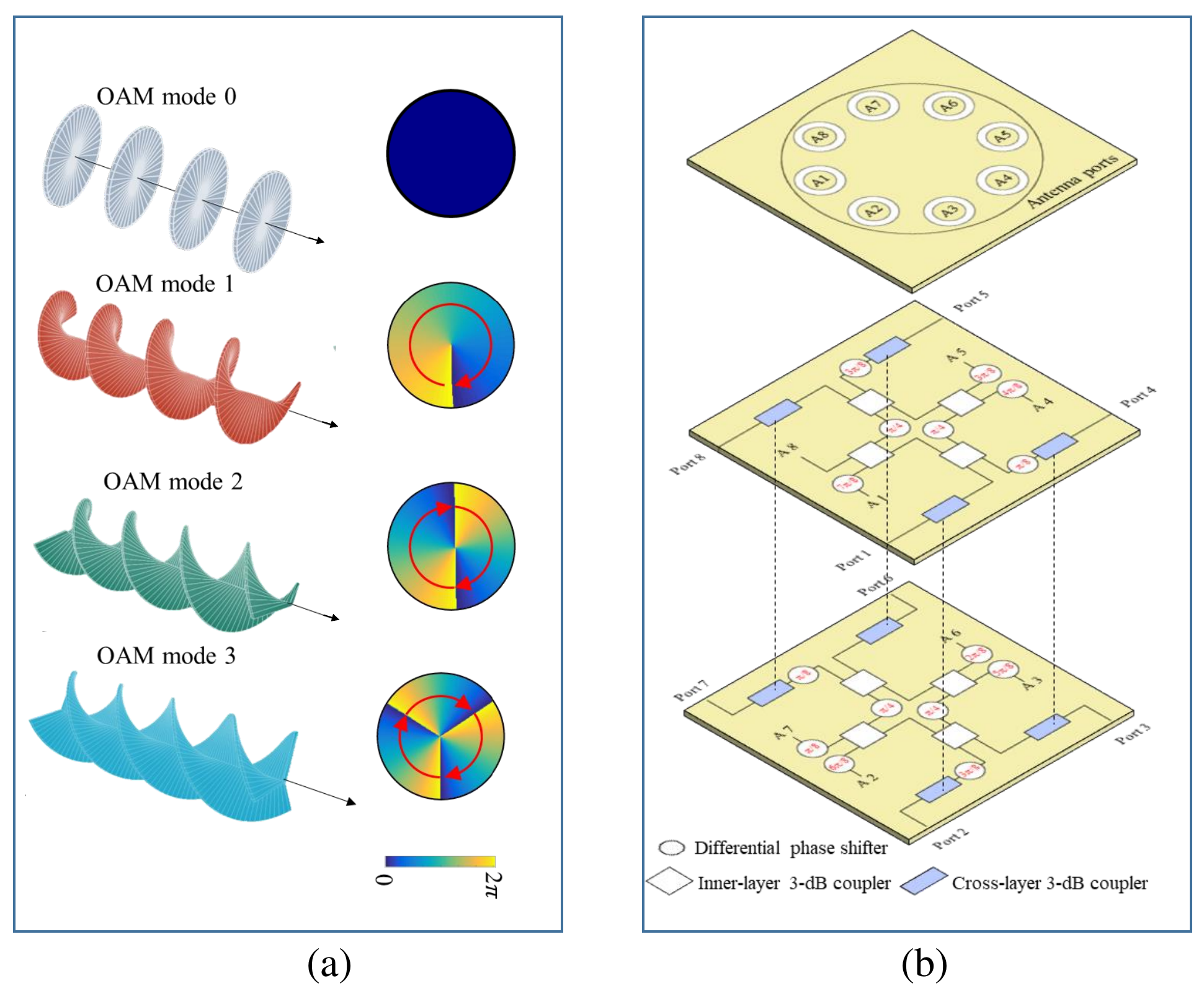}
    \end{center}
    \caption{An illustration of a) the wavefront of the OAM modes and spatial phase distribution on a plane vertical to the propagation axis and b) schematic design of multi-layer $8\times8$ Butler matrix for OAM multiplexing \cite{sas}. }
    \label{fig3}
\end{figure}

OAM multiplexing is based on the physical property of electromagnetic (EM) beams with spatial phase and amplitude distributions, where beams with different OAM modes appear orthogonal. A 1.58 Tbps wireless transmission has been demonstrated over a wideband  Butler matrix, forming a uniform circular array (UCA) in the sub-THz band. An illustration of the OAM modes and associated spatial phase distribution is given in Fig. \ref{fig3}(a), whereas Fig. \ref{fig3}(b) illustrates the schematic design of multi-layer 8x8 Butler matrix for OAM multiplexing. Specifically, the Butler matrix is a low-power passive analog circuit that generates and separates OAM beams without digital signal processing. The designed Butler matrix in Fig. \ref{fig3}(b) provides mode isolation of more than 15 dB in 135 to 170 GHz, allowing up to sixteen streams without harmful interference over the eight OAM modes and dual polarization.

\begin{figure}
    \begin{center}         \includegraphics[width=1\linewidth]{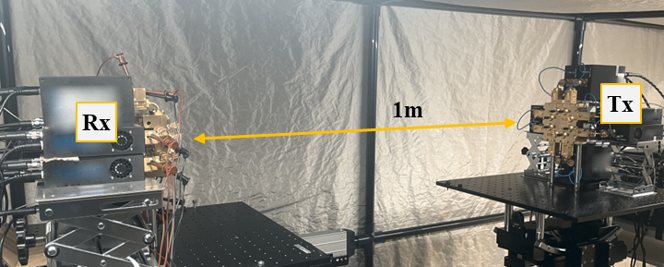}
    \end{center}
    \caption{Experimental setup in a shielded room \cite{sas}.}
    \label{fig4}
    %\vspace{-4mm}
\end{figure}

Fig. \ref{fig4} illustrates the experimental setup in the shield room, featuring the OAM multiplexing and embedded with the given Butler matrix, where the propagation axes of the oppositely arranged UCAs are aligned using optical lasers before transmission. All OAM waves are transmitted simultaneously. The diameter of the UCAs and transmission distance were 6 cm and 1 m, respectively. This distance can be extended arbitrarily by enlarging the diameter of UCAs or by using parabolic reflectors or dielectric lens to expand the effective diameter of the UCAs of the Butler matrix, and Yagi et al. reported a successful distance extension to 20 m using parabolic reflectors \cite{exp}.

%\vspace{-4mm}
\subsection{Future Research Directions and Visions}
As outlined above, implementing next-generation network faces several challenges, such as ensuring communication reliability over THz band, managing coexistence of low data rates and high-end applications, optimizing spectral and energy efficiency, and addressing privacy concerns. In light of these challenges, the following are some key future opportunities.

\textit{Energy/Spectral Efficient Adoption:}   
Energy efficiency is a key focus for IMT 2030, with ongoing efforts to develop sustainable and green technologies, along with energy-saving hardware and protocols. This includes exploring advanced network designs like self-organizing networks and network slicing. Additionally, IMT 2030 combines various technologies, requiring further research into security, global interoperability, and smooth communication across different networks and devices. Moreover, innovations like RedCap must be expanded to meet the needs of 6G and beyond, ensuring user-friendly systems with features like web-based case reports and real-time data validation.

\textit{Leveraging 5G Numerologies:} 6G focuses on both high-end and low-data-rate IoT applications, making it important to have a flexible network that can support different needs. To address this, 3GPP Release 18 and beyond highlight a key 5G New Radio feature called multiple numerologies. This allows the network to choose the best sub-carrier spacing (SCS) based on the user's needs, wireless conditions, and how quickly data needs to be sent.  However, the full potential of numerology selection is still unexplored, offering opportunities for further QoS-based customization to meet specific application needs with optimal service quality.

\textit{Seamless RIS Integration:} The dynamic nature of wireless channels continues to pose significant challenges, particularly in ensuring performance over THz waves. To address this, wireless solutions have increasingly focused on reconfiguring the environment through engineered surfaces. Known for its ability to control wireless signals, IRS is being considered a key technology for future wireless generations \cite{rohit}. However, IRS introduces substantial channel estimation overhead due to the large number of elements involved, presenting opportunities for research into low-complexity IRS solutions. Overall, addressing these challenges holds immense potential for transforming next-generation networks. IRS has already proven its utility in various innovations, including adaptive beamforming, data indexing, physical layer security, and  ubiquitous global connectivity \cite{ris}. Moreover, research efforts in this direction has been already initiated, e.g., the ETSI formed industry specification groups (ISG) to advance IRS research.

\textit{Advanced Security Solutions:} A significant challenge lies in ensuring data privacy, particularly in AI-based networks that rely on shared data. Innovative solutions are being explored to secure both the physical and upper-layer algorithms to address this. RIS has been suggested as a way to enhance physical layer security. Meanwhile, upper-layer encryption methods must be improved to handle the complexities of AI-powered networks. Interestingly, AI can strengthen and weaken security, where machine learning models can be optimized to detect threats, utilizing federated learning as a valuable approach, allowing models to allow devices without directly sharing data. 

%\vspace{-2mm}
\section{{Conclusion}}
{This article provided a comprehensive overview of the evolution of wireless communication technologies, from the foundations to recent advancements.} Then, we investigated the cutting-edge technologies in 5G and the ongoing progress towards 6G, reviewing the recommendations of IMT-2030. Further, we discussed specific features of IMT 2030, including three IMT-2020 extensions and three innovations. {In addition, we identified and analyzed three key challenges in implementing 6G, along with global standardization efforts.} Moreover, we demonstrated a proof of concept by transmitting THz signals using OAM multiplexing, a promising technology for THz transmission. Finally, we identified research opportunities and future directions inspired by IMT-2030 recommendations, paving the way for continued exploration and innovation in wireless communication.
%to inspire further potential research.

\bibliographystyle{IEEEtran}
%\bibliography{IEEEabrv,Ref}

\bibliography{IEEEabrv,BibRef}

\begin{thebibliography}{10}
\providecommand{\url}[1]{#1}
\csname url@samestyle\endcsname
\providecommand{\newblock}{\relax}
\providecommand{\bibinfo}[2]{#2}
\providecommand{\BIBentrySTDinterwordspacing}{\spaceskip=0pt\relax}
\providecommand{\BIBentryALTinterwordstretchfactor}{4}
\providecommand{\BIBentryALTinterwordspacing}{\spaceskip=\fontdimen2\font plus
\BIBentryALTinterwordstretchfactor\fontdimen3\font minus
  \fontdimen4\font\relax}
\providecommand{\BIBforeignlanguage}[2]{{%
\expandafter\ifx\csname l@#1\endcsname\relax
\typeout{** WARNING: IEEEtran.bst: No hyphenation pattern has been}%
\typeout{** loaded for the language `#1'. Using the pattern for}%
\typeout{** the default language instead.}%
\else
\language=\csname l@#1\endcsname
\fi
#2}}
\providecommand{\BIBdecl}{\relax}
\BIBdecl

\bibitem{imt}
R.~Liu, R.~Y.~N. Li, M.~Di~Renzo, and L.~Hanzo, ``A vision and an evolutionary
  framework for 6\uppercase{G}: Scenarios, capabilities and enablers,''
  \emph{arXiv preprint arXiv:2305.13887}, 2023.

\bibitem{thz}
A.~M. Elbir, K.~V. Mishra, S.~Chatzinotas, and M.~Bennis, ``Terahertz-band
  integrated sensing and communications: Challenges and opportunities,''
  \emph{IEEE Aerospace and Electronic Systems Magazine}, 2024.

\bibitem{book}
A.~F. Molisch, \emph{Wireless communications}.\hskip 1em plus 0.5em minus
  0.4em\relax John Wiley \& Sons, 2012, vol.~34.

\bibitem{red}
Z.~Shi and J.~Liu, ``A novel {NOMA}-enhanced {SDT} scheme for {NR} redcap in
  {5G/B5G} systems,'' \emph{IEEE Transactions on Wireless Communications},
  2023.

\bibitem{6gb}
S.~Kerboeuf, P.~Porambage, A.~Jain, P.~Rugeland, G.~Wikstr{\"o}m, M.~Ericson,
  D.~T. Bui, A.~Outtagarts, H.~Karvonen, P.~Alemany \emph{et~al.}, ``Design
  methodology for 6g end-to-end system: Hexa-x-ii perspective,'' \emph{IEEE
  Open Journal of the Communications Society}, 2024.

\bibitem{xmm}
Z.~Wang, J.~Zhang, H.~Du, E.~Wei, B.~Ai, D.~Niyato, and M.~Debbah, ``Extremely
  large-scale \uppercase{MIMO}: Fundamentals, challenges, solutions, and future
  directions,'' \emph{IEEE Wireless Communications}, vol.~31, no.~3, pp.
  117--124, Jun. 2024.

\bibitem{isc}
A.~Kaushik, R.~Singh, M.~Li, H.~Luo, S.~Dayarathna, R.~Senanayake, X.~An, R.~A.
  Stirling-Gallacher, W.~Shin, and M.~Di~Renzo, ``Integrated sensing and
  communications for {I}o{T}: Synergies with key {6G} technology enablers,''
  \emph{IEEE Internet of Things Magazine}, vol.~7, no.~5, pp. 136--143, 2024.

\bibitem{ubc}
H.~Lee, B.~Lee, H.~Yang, J.~Kim, S.~Kim, W.~Shin, B.~Shim, and H.~V. Poor,
  ``Towards 6\uppercase{G} hyper-connectivity: Vision, challenges, and key
  enabling technologies,'' \emph{Journal of Communications and Networks},
  vol.~25, no.~3, pp. 344--354, Jun. 2023.

\bibitem{hrl}
A.~Kaushik, R.~Singh, S.~Dayarathna, R.~Senanayake, M.~Di~Renzo, M.~Dajer,
  H.~Ji, Y.~Kim, V.~Sciancalepore, A.~Zappone, and W.~Shin, ``Toward integrated
  sensing and communications for 6{G}: Key enabling technologies,
  standardization, and challenges,'' \emph{IEEE Communications Standards
  Magazine}, vol.~8, no.~2, pp. 52--59, Jun. 2024.

\bibitem{aic}
Z.~Chen, Z.~Zhang, and Z.~Yang, ``Big {AI} models for {6G} wireless networks:
  Opportunities, challenges, and research directions,'' \emph{IEEE Wireless
  Communications}, 2024.

\bibitem{quant}
R.~Singh and R.~M. Bodile, ``A quick guide to quantum communication,''
  \emph{arXiv preprint arXiv:2402.15707}, 2024.

\bibitem{sas}
H.~Sasaki, H.~Yagi, R.~Kudo, and D.~Lee, ``1.58 \uppercase{T}bps {OAM}
  multiplexing wireless transmission with wideband {B}utler matrix for
  sub-{THz} band,'' \emph{IEEE J. Sel. Areas Commun}, vol.~42, no.~6, pp.
  1613--1625, Jun. 2024.

\bibitem{exp}
Y.~Yagi, H.~Sasaki, and D.~Lee, ``Parabolic reflector for \uppercase{UCA}-based
  \uppercase{OAM} multiplexing in sub-\uppercase{T}hz band and transmission
  experiment,'' in \emph{IEEE Workshop on High Capacity Wireless Communications
  (HCWC 2023)}, Dec. 2023.

\bibitem{rohit}
R.~Singh, A.~Kaushik, W.~Shin, G.~C. Alexandropoulos, M.~Toka, and M.~Di~Renzo,
  ``Indexed multiple access with reconfigurable intelligent surfaces: The
  reflection tuning potential,'' \emph{IEEE Communications Magazine}, vol.~62,
  no.~4, pp. 120--126, Apr. 2024.

\bibitem{ris}
M.~Toka, B.~Lee, J.~Seong, A.~Kaushik, J.~Lee, J.~Lee, N.~Lee, W.~Shin, and
  H.~V. Poor, ``{RIS}-empowered {LEO} satellite networks for {6G}: {P}romising
  usage scenarios and future directions,'' \emph{IEEE Communications Magazine},
  vol.~62, no.~11, pp. 128--135, Nov. 2024.

\end{thebibliography}

%\vspace{-2mm}

\end{document}